\documentclass[pre,superscriptaddress,twocolumn,footinbib]{revtex4-1}
\pdfoutput=1
\usepackage[colorlinks=true,urlcolor=blue]{hyperref}
\usepackage[normalem]{ulem}
\usepackage{mathtools}
\usepackage{amsfonts}
\usepackage{graphicx}
\usepackage{stmaryrd}
\usepackage{amsmath}
\usepackage{amssymb}
\usepackage{lineno}
\usepackage{xcolor}
\usepackage{color}
\usepackage{bbold}
\usepackage{bm}

\renewcommand{\Re}{\mathfrak{R}}

\DeclareMathOperator{\De}{De}
\DeclareMathOperator{\tr}{tr}

\newcommand{\traceless}[1]{\left\llbracket #1 \right\rrbracket}

\definecolor{red}{rgb}{0.75,0,0}
\definecolor{blue}{rgb}{0,0,0.75}
\definecolor{green}{rgb}{0,0.5,0}

\newcommand\thefont{\expandafter\string\the\font}

\makeatother

\begin{document}
	
\title{Hydrodynamic stability and pattern formation in hexatic epithelial layers}

\author{Josep-Maria Armengol-Collado}
\affiliation{Instituut-Lorentz, Universiteit Leiden, P.O. Box 9506, 2300 RA Leiden, The Netherlands}

\author{Leonardo Puggioni}
\affiliation{Instituut-Lorentz, Universiteit Leiden, P.O. Box 9506, 2300 RA Leiden, The Netherlands}

\author{Livio N. Carenza}
\affiliation{Instituut-Lorentz, Universiteit Leiden, P.O. Box 9506, 2300 RA Leiden, The Netherlands}
\affiliation{Physics Department, College of Sciences, Ko{\c c} University, Rumelifeneri Yolu 34450 Sar\i{}yer, Istanbul,  T{\" u}rkiye}

\author{Luca Giomi}
\email{giomi@lorentz.leidenuniv.nl}
\affiliation{Instituut-Lorentz, Universiteit Leiden, P.O. Box 9506, 2300 RA Leiden, The Netherlands}

\begin{abstract}
We investigate the hydrodynamic stability and the formation of patterns in a continuum model of epithelial layers, able to account for the interplay between mechanical activity, lateral adhesion and the $6-$fold orientational order originating from the hexagonal morphology of the cells. Unlike in other models of active liquid crystals, the balance between energy injection and dissipation can here involve multiple length scales, resulting in a large spectrum of dynamical behaviors. When kinetic energy is dissipated by the cells' adhesive interactions at a rate higher that at which is injected by active stresses, the quiescent state of the cellular layer is {\em generically stable}: i.e. hydrodynamically stable regardless of its size. On the other hand, as the cellular layer becomes progressively more active, this homeostatic condition is altered by a hierarchy of pattern-forming instabilities, where the system organizes in an increasingly large number of counter-flowing lanes of fixed width. In two-dimensional periodic domains, the latter organization is itself unstable to the proliferation of vortices and the dynamics of the cellular layer becomes eventually chaotic and yet different from the more common {\em active turbulence}.
\end{abstract}

\maketitle

\section{Introduction}

Among the avenues to a continuum mechanics of multicellular systems, liquid crystals hydrodynamics is one the most actively  pursued. In addition to the great deal of physical intuition and theoretical machinery already at hand, the rich phenomenology of flowing liquid crystals provides {\em de facto} a paradigm for rationalizing the interplay between spatial organization and collective motion: a pivotal mechanism in living matter. Examples include the dynamics of topological defects in confluent cell layers~\cite{Duclos:2017,Kawaguchi:2017,Saw:2017,Balasubramaniam:2021,Yashunsky:2022,Hoffmann:2022} and their possible role in tissue morphogenesis~\cite{Maroudas:2021}, cancer progression~\cite{Ilina:2020} and wound healing~\cite{Tetley:2019}, as well as a vast spectrum of pattern-forming instabilities leading to migratory behaviors~\cite{Ranft:2010,Duclos:2018,Czajkowski:2018,Hoffmann:2020,Grossman:2022}. Much of these models, in turn, revolves around the notion of polar or nematic order: i.e. the property of elongated building blocks, such as bacteria or mesenchymal cells, to form bulk phases with local 1- or 2-fold rotational symmetry respectively. The central idea behind this approach is that uniaxial alignment focuses the microscopic forces actively generated by the cells, thus giving rise to contractile or extensile stresses at the mesoscopic length scale. The latter, then, conspire with other forces, such as those originating from the frictional interactions of the cells between each other and with a substrate, thereby leading to collective motion.

Migratory epithelia differ from the above in that orientational order is not strictly uniaxial, thus eluding the standard classification in terms on polar (i.e. 1-fold) or nematic (i.e. 2-fold) liquid crystals. An alternative strategy, built upon preliminary evidence from several cell-resolved models of epithelia~\cite{Li:2018,Durand:2019,Pasupalak:2020,Li:2021}, was proposed in Refs.~\cite{Armengol:2023,Armengol:2024} and revolves around the notion of {\em multiscale order} (MsO): i.e. the coexistence of different ordered structures at different length scales. While this hierarchical form of organization is not limited to orientational order~\cite{Torquato:2018,Torquato:2019} nor to cellular matter~\cite{Formanek:2023}, the notion of epithelia as multiscale active liquid crystals has naturally emerged as a possible framework to relate the spatial structure and the mechanical behavior of a class of active materials whose organization is too complex to be described in terms of a single order parameter. 

The simplest continuum model of epithelial layer based on MsO comprises only two {\em layers} of orientational order, that is hexatic and nematic~\cite{Armengol:2024}. The former stems from the local 6-fold symmetry resulting from the close-packing of naturally isotropic, but deformable, building blocks, while the latter reflects the typical uniaxial organization of collectively migrating {\em cell clusters}~\cite{Armengol:2023}. It follows that hexatic order is dominant at small length scales, while the large scale behavior is mainly determined by the nematic order emerging among cell clusters, with the average cluster size marking the length scale at which such a {\em hexanematic crossover} take place~\cite{Eckert:2023}. 

Now, because of its historical importance and its connection with active polar- and nemato-hydrodynamics, much of the literature on flowing epithelia has so far been limited to large scale collective phenomena, where hexatic order is neglible or nearly so. These include the chaotic motion observed in cultures of spindle-like cells (e.g. HT1080, RPE1 etc.), both in open domains \cite{Duclos:2017,Blanch:2018}, under confinement~\cite{Duclos:2018,Yashunsky:2022}, or {\em in vivo}~\cite{Chepizhko:2024}, the formation of vortices in freeley expanding epithelia~\cite{Heinrich:2020}, as well as various other examples of pattern formation involving integer and semi-integer topological defect~\cite{Duclos:2017,Saw:2017,Blanch:2018,Balasubramaniam:2021,Hoffmann:2022} (see also Ref.~\cite{Saw:2018} for an overview). By contrast, our understanding of the role of hexatic order in epithelia has only recently started to bloosom and the questions still outnumber the answers by far. In Ref.~\cite{Krommydas:2023}, for instance, some of the authors showed how the unbinding of $\pm 1/6$ hexatic defects can account for the small scale dynamics observed in epithelia during cell intercalation: i.e. the process that allows confluent cells to move by remodeling the network of intercellular junctions in their surrounding. The interplay between hexatic order and cell division has also emerged as a potentially relevant mechanism during morphogenesis~\cite{Tang:2023}, while the connection between the onset of collective migration and the Kosterlitz-Thouless-Halperin-Nelson-Young (KTHNY) melting scenario has been now convincingly demonstrated by different cell-resolved models~\cite{Li:2018,Durand:2019,Pasupalak:2020,Li:2021}, although its implications in the spatiotemporal organization of epithelia is still entirely elusive.     
 
In this article, we provide the first exhaustive analysis of the hydrodynamic stability of hexatic epithelia, both under confinement and in periodic domains. After a brief overview of hexatic hydrodynamics (Sec.~\ref{sec:hydrodynamiceq}), we focus on energy dissipation and show how the small scale hydrodynamic stability of our hexatic epithelia crucially demands a scale-dependent balance between energy injection and dissipation mechanisms, with the latter naturally arising from the cells lateral interactions (Sec.~\ref{sec:regularization}). To illustrate this concept, we specifically investigate two different realizations of epithelia under channel confinement and demonstrate the existence of a variety of instabilities leading to collective migration (Sec.~\ref{sec:periodic_boundary}). While the simplest of these recapitulates all the essential aspects of the classic spontaneous flow transition of active liquid crystals~\cite{Voituriez:2005,Edwards:2009}, the availability of multiple inherent length scales, originating from the scale-dependent balance between energy injection and dissipation, results into a cascade of further instabilities. These allow cells to partition the confining channels in an arbitrary number of oppositely flowing lanes. In Sec.~\ref{sec:2D} we show how much of the intuition developed for the quasi-one-dimensional flows spontaneously arising under channel confinement carries over to two-dimensional periodic flows and anticipate the emergence of a new class of ``active turbulence'', whose spatial organization is reminiscent of the herringbone patterns observed in certain kind of malignant mesenchymal tumors such as fibrosarcoma~\cite{McKee:2020}.

\section{\label{sec:hydrodynamiceq}Small scale hydrodynamics of hexatic epithelia}

\begin{figure}
	\centering
	\includegraphics[width=\columnwidth]{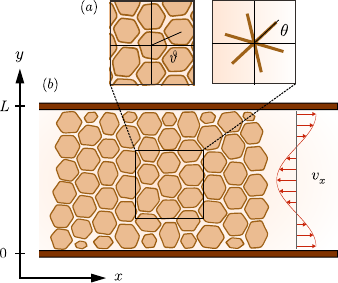}
	\caption{\label{fig:schematic}
		(a) Schematic representation of cells with $6$-fold orientational symmetry (left) together with the coarse-grained hexatic director (right). (b) Illustration of a two-dimensional channel that is infinite in the $x$-direction and has width $L$ in the $y$-direction.}
\end{figure}

The hydrodynamic equations of an epithelial layer featuring multiscale {\em hexanematic} order have been derived in Ref.~\cite{Armengol:2024} upon extending the hydrodynamic theory of passive $p-$atic liquid crystals~\cite{Giomi:2021a, Giomi:2021b}. A central quantity in this formulation is the $p-$atic order parameter tensor $\bm{Q}_{p}=Q_{i_{1}i_{2}\cdots\,i_{p}}\bm{e}_{i_{1}}\otimes \bm{e}_{i_{2}}\cdots\,\bm{e}_{i_{p}}$, with $i_{n}\in\{x,y\}$ and $n=1,\,2\ldots\,p$, expressing the local {\em average} configuration of the $p-$atic mesogens. This is given by
\begin{equation}
\bm{Q}_{p} = \sqrt{2^{p-2}}\,\traceless{\left<\bm{\nu}^{\otimes p}\right>}=\sqrt{2^{p-2}}\,|\Psi_p|\traceless{\bm{n}^{\otimes p}}\;,
\end{equation}
where $\bm{\nu} = \cos{\vartheta}\,\bm{e}_x + \sin{\vartheta}\,\bm{e}_y$ is the orientation of the individual building blocks, as depicted in Fig.~\ref{fig:schematic}a, $\langle \cdots \rangle$ the ensemble average and the operator $\traceless{\dots}$ has the effect of rendering its argument to be symmetric and traceless. The quantity $|\Psi_p|$ is the magnitude of the complex $p-$atic order parameter and, together with its phase $\theta$, is obtained upon taking the ensemble average of the $p-$fold orientation field $\psi_{p}=e^{ip\vartheta}$, i.e. $\Psi_{p}=\langle \psi_{p} \rangle = |\Psi_{p}|e^{ip\theta}$. The unit vector, $\bm{n} = \cos{\theta}\,\bm{e}_x + \sin{\theta}\,\bm{e}_y$, finally, is the $p-$atic director, marking the average orientation at the scale of a volume element (see Fig.~\ref{fig:schematic}b). By construction, $|\bm{Q}_{p}|^{2}=Q_{i_{1}i_{2}\cdots\,i_{p}}Q_{i_{1}i_{2}\cdots\,i_{p}}=|\Psi_{p}|^{2}/2$.

In the following, we assume an epithelial layer that is confined at length scales well below the hexanematic crossover scale, so that its dynamics results solely from the interplay between hexatic order and the contractile or extensile forces generated at the cellular scale, while nematic order is assumed negligible. The hydrodynamic equations are then given in terms of the mass density $\rho$, momentum density $\rho\bm{v}$ and the rank$-6$ order parameter tensor $\bm{Q}_{6}$, by
\begin{subequations}\label{eq:hydrodynamics}
\begin{gather}
\frac{D\rho}{Dt}+\rho\nabla\cdot\bm{v} = (k_{\rm d}-k_{\rm a})\rho\;,\\
\rho\,\frac{D\bm{v}}{Dt}=\nabla\cdot\left(\bm{\sigma}^{({\rm p})}+\bm{\sigma}^{({\rm a})}\right)+\bm{f}\;,\\
\frac{D\bm{Q}_6}{Dt}=\Gamma_{6}\bm{H}_6+6\traceless{\bm{Q}_6\cdot\bm{\omega}}+\lambda_{6}\traceless{\nabla^{\otimes 4}\bm{u}} \notag \\
+\overline{\lambda}_{6}\tr(\bm{u})\bm{Q}_{6}+\nu_{6}\traceless{\bm{u}^{\otimes3}}\;,
\end{gather}
\end{subequations}
with $D/Dt = \partial_{t}+\bm{v}\cdot\nabla$ the material derivative. In Eq.~(\ref{eq:hydrodynamics}a), $k_{\rm d}$ and $k_{\rm a}$ are the rates of cell division and apoptosis. In Eq.~(\ref{eq:hydrodynamics}b), the tensors $\bm{\sigma}^{({\rm p})}$ and $\bm{\sigma}^{({\rm a})}$ embody the passive and active stresses driving the cellular flow, whereas $\bm{f}$ is an external body force. In Eq.~(\ref{eq:hydrodynamics}c), $\Gamma_{6}^{-1}$ is the rotational viscosity, $\bm{u}=[(\nabla\bm{v})+(\nabla\bm{v})^{\intercal}]/2$ and $\bm{\omega}=[(\nabla\bm{v})-(\nabla\bm{v})^{\intercal}]/2$, with $\intercal$ indicating transposition, are the strain rate and vorticity tensor expressing the feedback of the flow on the configuration of the hexatic order parameter, with $\lambda_{6}$, $\overline{\lambda}_{6}$ and $\nu_{6}$ phenomenological constants and $(\nabla^{\otimes n})_{i_{1}i_{2}\cdots\,i_{p}}=\partial_{i_{1}}\partial_{i_{2}}\ldots\,\partial_{i_{n}}$. The rank$-6$ tensor $\bm{H}_{6}=-\delta F/\delta\bm{Q}_{6}$ is the analog of the molecular tensor in nematic liquid crystals and drives the relaxational dynamics towards the minimum of the free energy
\begin{equation}
F = \int {\rm d}^{2}r\,\left(\frac{L_{6}}{2}\,|\nabla\bm{Q}_{6}|^{2}+\frac{A_{6}}{2}\,|\bm{Q}_{6}|^{2}+\frac{B_6}{4}\,|\bm{Q}_{6}|^{2}\right)\;.
\end{equation}
Here $L_{6}$ is the order parameter stiffness and $A_{6}$ and $B_{6}$ phenomenological constants setting the magnitude of the order parameter at equilibrium: i.e. $|\Psi_{6}^{(0)}|=\sqrt{-2A_{6}/B_{6}}$. The passive stress tensor in Eq.~(\ref{eq:hydrodynamics}b) is routinely decomposed as $\bm{\sigma}^{({\rm p})}=-P\mathbb{1}+\bm{\sigma}^{({\rm e})}+\bm{\sigma}^{({\rm v})}+\bm{\sigma}^{({\rm r})}$, where $P$ is the pressure and $\sigma_{ij}^{({\rm e})}=- L_{6}\,\partial_{i}\bm{Q}_{6}\odot\partial_{j}\bm{Q}_{6}$, with $\odot$ indicating a contraction of all the matching indices of the two operands, the elastic stress arising in response to a static deformation of a fluid patch. Lastly, the tensors $\bm{\sigma}^{({\rm r})}=-\overline{\lambda}_{6}\,\bm{Q}_{6}\odot\bm{H}_{6}\,\mathbb{1}-\lambda_{6}\nabla^{\otimes 4}\odot\bm{H}_{6}+3(\bm{Q}_{6}\cdot\bm{H}_{6}-\bm{H}_{6}\cdot\bm{Q}_{6})$ and $\bm{\sigma}^{({\rm v})}$ convey the {\em reactive} (i.e. energy preserving) and {\em dissipative} (i.e. entropy producing) stresses originating from the interplay between hexatic order and flow. 

The active stress $\bm{\sigma}^{({\rm a})}$ was constructed in Ref.~\cite{Armengol:2024} on the basis of phenomenological and microscopic arguments and, in the specific case of active hexatics, takes the form
\begin{equation}\label{eq:activestress}
\bm{\sigma}^{({\rm a})}=\alpha_{6}\,\nabla^{\otimes 4}\odot\bm{Q}_{6}\;,
\end{equation}
where the activity coefficient $\alpha_{6} \sim -\mathcal{F}a^{5}$ reflects the force per unit area, $\mathcal{F}$, arising from the tension imbalance at the vertices of the cellular network at the length scale of the typical cell size $a$. Although the microscopic tension is mainly contractile, the resulting force density can be either contractile (i.e. $\mathcal{F}<0$ and $\alpha_{6}>0$) or extensile (i.e. $\mathcal{F}>0$ and $\alpha_{6}<0$) depending on the specific tension distribution and cellular geometry~\cite{Armengol:2024}.

\section{\label{sec:regularization}Dimensional reduction and hyperviscous regularization}

The main goal of this study is to investigate the hydrodynamic stability of a hexatically ordered epithelial layer confined at length scales where nematic order is instead negligible. To this end, we assume the system enclosed in a two-dimensional channel, whose length is much larger than the width $L$ (see Fig.~\ref{fig:schematic}), so to render the configuration of the cell layer invariant under translations along the longitudinal direction, say $\bm{e}_{x}$. Furthermore, we assume the system at homeostasis, so that $k_{\rm d}=k_{\rm a}$, and homogeneous, from which $\rho={\rm const}$ and $\nabla\cdot\bm{v}=\partial_{y}v_{y}=0$. Thus $\theta=\theta(y)$, $\bm{v}=v_{x}(y)\bm{e}_{x}$, having taken into account that $v_{y}(0)=v_{y}(L)=0$ because of the lateral confinement. To further isolate the interplay between energy injection and dissipation, which serves as running engine of collective migration, we simplify our treatment of flow alignment by retaining only the corotational term in Eq.~(\ref{eq:hydrodynamics}c) and ignore passive backflow effects, being the latter unimportant in athermal systems. Both simplifications only affect the details of the problem, without changing the general picture. Under these assumptions, Eqs.~\eqref{eq:hydrodynamics} reduce to the following coupled equations for the angle $\theta$ and the longitudinal velocity $v_{x}$:
\begin{subequations}\label{eq:generalsysteminstability}
\begin{gather}
\partial_t \theta=\mathcal{D}_{6}\partial_y^2\theta-\frac{1}{2}\,\partial_yv_x\;,\label{eq:generalsysteminstability1}\\
\rho\,\partial_t v_x = \partial _y \left(\sigma_{xy}^{({\rm v})}+\frac{1}{8}\,\alpha_{6}\partial_y^4\sin{6\theta}\right)\;,\label{eq:generalsysteminstability2}
\end{gather}	
\end{subequations}
where $\mathcal{D}_{6}=\Gamma_{6}L_{6}$ is the rotational diffusion coefficient of the hexatic phase. 

Eqs.~\eqref{eq:generalsysteminstability} provide a {\em minimal} continuum model of collective epithelial migration under strong confinement and are now amenable to a thorough analytical investigation. In a regime of steady laminar flow (i.e. Stokesian regime), Eqs.~\eqref{eq:generalsysteminstability} imply
\begin{subequations}\label{eq:constantshearstress}
\begin{gather}
v_{x} = v_{0} + 2\mathcal{D}_{6}\partial_{y}\theta\;,\\	
\sigma_{0} = \sigma_{xy}^{({\rm v})}+\frac{1}{8}\,\alpha_{6}\partial_y^4\sin{6\theta}\;,
\end{gather}
\end{subequations}
with $v_{0}$ and $\sigma_{0}$ integration constants, whose specific expression is determined by the boundary conditions. Eqs.~(\ref{eq:constantshearstress}) demand the collective cellular flow to be directed by the pattern of spatial variations of the cells' average orientation, while the stress delivered by the actin cytoskeleton is {\em instantaneously} dissipated by viscosity. This, in turn, originates from the cell lateral interactions and, as suggested by recent numerical studies of the Vertex model~\cite{Hertaeg:2022,Tong:2022}, can lead to a rich landscape of exotic rheological phenomena. 

\begin{figure}
\centering
\includegraphics[width=\columnwidth]{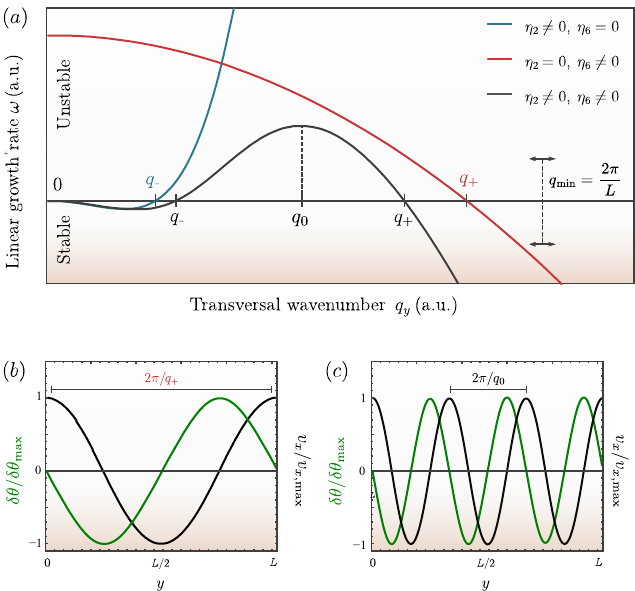}
\caption{\label{fig:dispersion} \textbf{One-dimensional dispersion relation and stationary states} (a) Graphical representation of the dispersion relation deduced in Eq.~\eqref{eq:groutheta0eta1eta2} for three different cases of $\bm{\sigma}^{({\rm v})}$: $\eta_2\neq0$, $\eta_4=\eta_6=0$ (blue curve), $\eta_2=\eta_4=0$, $\eta_6\neq0$ (red curve) and $\eta_4=0$, $\eta_2\neq0$ $\eta_6\neq0$ (black curve). The label $q_{\min}=2\pi/L$ marks the lowest wave number available in the system. Upon increasing the size $L$ of the channel, $q_{\min}$ moves toward smaller and smaller values, until the quiescent state becomes unstable when $q_{\min}=q_{+}$. (b) and (c) Examples of the stationary configuration of the local hexatic orientation $\delta\theta$ (green) and velocity $v_{x}$ (black) when (b) $\eta_2=0$, $\eta_6\neq0$ and (c) $\eta_2\neq0$, $\eta_6\neq0$.}
\end{figure}

Crucially, Eqs.~\eqref{eq:generalsysteminstability} entail intrinsic length scales, originating from the dynamical equilibrium between energy injection and dissipation at the rate imposed by the relaxation of the internal structure of the epithelial layer, here embodied in the hexatic orientation $\theta$. To illustrate this concept, let us first consider the standard linear constitutive equation relating shear stress and rate, that is
\begin{equation}\label{eq:viscousstresseta0}
\sigma_{xy}^{({\rm v})} = \eta_{2}\partial_{y}v_{x}\;,
\end{equation}
being $\eta_{2}$ the shear viscosity. In this case, Eqs.~\eqref{eq:generalsysteminstability} always admit the trivial solution $\theta(y,t)=\theta_0$, $v_x(y,t)=v_0$, with $\theta_{0}$ a constant. To assess the stability of this solution with respect to an infinitesimal perturbation, we substitute Eqs.~(\ref{eq:constantshearstress}b) and \eqref{eq:viscousstresseta0} in Eq.~(\ref{eq:constantshearstress}a), to obtain
\begin{equation}\label{eq:dthetadt}
\partial_{t}\theta=\mathcal{D}_{6}\partial_{y}^{2}\theta+\frac{\alpha_{6}}{16\eta_{2}}\,\partial_{y}^{4}\sin 6\theta\;.
\end{equation}
Next, we decompose $\theta(y,t)=\theta_0+\delta\theta(y,t)$ and express the perturbations in terms of Fourier-Laplace components: i.e. $\delta\theta(y,t)=\sum_{n=-\infty}^{\infty} \int {\rm d}\omega\,\theta(q_n,\omega)\,e^{iq_{n}y+\omega t}$, with $q_{n}=2\pi n/L$ and $\omega\in\mathbb{C}$~\cite{Cross:1993}. In the following, we will omit the index $n$ to avoid clutter. Substituting this in Eq.~\eqref{eq:dthetadt} yields 
\begin{equation}\label{eq:dispersion}
\omega =-\mathcal{D}_{6}q^2+\frac{3\alpha_{6}\cos{6\theta_0}}{8\eta_{2}}\,q^4;.
\end{equation}
Now, if $\alpha_{6}=0$, we recover the standard diffusive mode, for which $\omega = -\mathcal{D}_{6}q^{2}$. As $\mathcal{D}_{6}>0$, $\Re\{\omega\}<0$ and the amplitude of an arbitrary infinitesimal perturbation vanishes at short length scales, that is in the limit $q\to\infty$. In the presence of hexatic activity, on the other hand, $\omega \sim \alpha_{6}\,q^4$ for large $q$ values and the same behavior is no longer guaranteed. Specifically, given an initial orientation $\theta_0$, there is always a range of activities for which a perturbation of sufficiently large wave number will give rise to a unstable branch (see blue curve in Fig.~\ref{fig:dispersion}a). For instance, in the presence of contractile activity (i.e. $\alpha_{6}>0$) and for $\theta_0=0$, the dispersion relation given by Eq.~\eqref{eq:dispersion} demands the homogeneous state of a confined epithelial layer to be unstable to perturbations whose wave number is larger than
\begin{equation}\label{eq:ellmin}
q_{-} = \sqrt{\frac{8\eta_{2}\mathcal{D}_{6}}{3\alpha_{6}}}\;.
\end{equation}
Such a scenario is obviously unphysical as it would imply instability already at the cellular scale, thereby preventing the formation of structures of any kind. Accounting for passive backflow effects, via the reactive stress $\sigma_{xy}^{({\rm r})}=9L_{6}|\Psi_{6}|^{2}\partial_{y}^{2}\theta$, contributes to the numerator of Eq.~\eqref{eq:dispersion} with an additional quadratic term, but this does not influence the structure of hydrodynamic modes at neither large nor small length scales.

Now, the standard approach towards regularizing this behavior consists of manipulating the energy dissipation rate across length scales by including in the expression of the viscous stress, Eq.~\eqref{eq:viscousstresseta0}, higher order velocity gradients. This strategy, which has been previous used to investigate the behavior of active suspensions \cite{Slomka:2015} and as a possible source of turbulence in self-driven systems \cite{Linkmann:2019}, amounts to the following expression of the viscous stress:
\begin{equation}\label{viscousstressfull}
\bm{\sigma}^{({\rm v})}=2\sum_{k=1}^{k_{\max}}\left(-1\right)^{k-1}\eta_{2k}\nabla^{2(k-1)}\traceless{\bm{u}}\;,
\end{equation}
where $\eta_{2k}$ is a generalized viscosity coefficient, often referred to as {\em hyperviscosity}, and $\nabla^{2(k-1)}$ stands for the $(k-1)-$th recursive application of the Laplace operator $\nabla^{2}$. On the other hand, the spectral density $\mathcal{E}=\mathcal{E}(q)$ of the kinetic energy stored in a unit mass -- i.e. $\langle v^{2}/2 \rangle = \int {\rm d}q\,\mathcal{E}(q)$ -- changes in time at the rate 
\begin{equation}\label{eq:dissipation_rate}
\frac{{\rm d}\mathcal{E}(q)}{{\rm d}t} = \left\langle \mathcal{F}_{-\bm{q}}\left\{\bm{v}\right\}\cdot\mathcal{F}_{\bm{q}}\left\{\nabla\cdot\left(-\bm{v}\otimes\bm{v}+\rho^{-1}\nabla\cdot\bm{\sigma}\right)\right\}\right\rangle\;,
\end{equation}
where $\langle \cdots \rangle$ denotes a statistical average and $\mathcal{F}_{\bm{q}}\{\cdots\}=\int {\rm d}^{2}r\,e^{-i\bm{q}\cdot\bm{r}}(\cdots)$ Fourier transformation (see e.g. Refs.~\cite{Alexakis:2018,Carenza:2020a,Carenza:2020b}). Taken together, Eqs.~\eqref{viscousstressfull} and \eqref{eq:dissipation_rate} implies ${\rm d}\mathcal{E}(q)/{\rm d}t \sim  -\nu_{2}q^{2}-\nu_{4}q^{4}-\nu_{6}q^{6}+\cdots$, with $\nu_{2k}=\eta_{2k}/\rho$ is the kinematic hyperviscosity. This scaling behavior, on the other hand, must reflect that of energy injection in order for the system to be thermodynamically stable and avoid the accumulation of energy at a given scale or range of scales. The latter provides a criterion for the truncation of the expansion in Eq.~\eqref{viscousstressfull}. That is
\begin{equation}
k_{\max}\geq p\;,
\end{equation}
for generic active $p-$atic liquid crystals. Thus, for the specific case of $p=6$, Eq.~\eqref{viscousstressfull} can be cast in the form
\begin{equation}
\bm{\sigma}^{({\rm v})}=2\left(\eta_{2}-\eta_{4}\nabla^{2}+\eta_{6}\nabla^{4}\right)\traceless{\bm{u}}\;.
\end{equation}
Notice that, while $\eta_{4}$ could be either positive or negative, indicating a sink or source of energy respectively, $\eta_{6}>0$ for thermodynamic stability. Including those corrections in the momentum balance relation given in Eq.~\eqref{eq:constantshearstress} yields
\begin{equation}\label{completeviscousstress}
\sigma_{xy} = \eta_2  \partial_yv_x-\eta_4  \partial_y^3v_x+\frac{1}{8}\,\alpha_{6}\partial_y^4\,\sin{6\theta}+\eta_6  \partial_y^5v_x\;,
\end{equation}
from which one can again derive a dispersion relation in the form
\begin{equation}\label{eq:groutheta0eta1eta2}
\omega =-\mathcal{D}_{6} q^2 + \frac{3\alpha _6 \cos{6\theta_0}\,q^4}{8\left(\eta_2+\eta_4 q^2+\eta _6 q^4\right)}\;.
\end{equation}
In the $q\to\infty$ limit, the second term on the right-hand side of this equation converges to a constant, indicating that energy is now dissipated at the same rate at which it is injected, whereas the first term guarantees hydrodynamic stability at small length scales, as $\omega\to-\infty$ for $\mathcal{D}_{6}>0$. 

Finally, extending this approach to generic active $p-$atics, for which~\cite{Armengol:2024}
\begin{equation}
\bm{\sigma}^{({\rm a})}	= \alpha_{p}\nabla^{\otimes(p-2)}\odot\bm{Q}_{p}\;,
\end{equation}
highlights the existence of two fundamental time scales, associated with the relaxation of linear momentum (i.e. $\tau_{v}$) and $p-$atic order (i.e. $\tau_{\theta}$) respectively. These are given by
\begin{equation}\label{eq:time_scales}
\tau_{v} = \frac{\rho\ell^{p}}{\eta_{p}}\;,\qquad
\tau_{\theta} = \frac{\ell^{2}}{\mathcal{D}_{p}}\;,
\end{equation}
with $\ell$ the characteristic length scale of the flow. Multiplying these for the shear rate $\dot{\epsilon}$ yields, in turn, the classic Reynolds number, ${\rm Re}=\dot{\epsilon}\tau_{v}$, expressing the ratio between the magnitude of inertial and viscous forces, and the Deborah number, ${\rm De}=\dot{\epsilon}\tau_{\theta}$, quantifying the rate of the flow in units of the inherent relaxation rate of the fluid~\cite{Reiner:1964}. In the case of a laminar active $p-$atic flow, the same reasoning behind Eq.~\eqref{eq:groutheta0eta1eta2} implies 
\begin{equation}
\dot{\epsilon}\sim\frac{\alpha_{p}}{\eta_{p}}\;,
\end{equation}
from which 
\begin{equation}\label{eq:re_and_de}
{\rm Re}=\frac{\rho|\alpha_{p}|\ell^{p}}{\eta_{p}^{2}}\;,\qquad
\De=\frac{|\alpha_{p}|\ell^{2}}{\eta_{p}\mathcal{D}_{p}}\;.
\end{equation}
In the following, we will assume that ${\rm Re}\ll 1$ at any length scale, as typical of cellular motion, and will show how $\De$ determines the hydrodynamic stability of the quiescent state, as well as the structure of the migratory flow resulting from the persistent injection and dissipation of energy.

\section{\label{sec:periodic_boundary}Pattern formation in periodic domains}

Hereafter, we will assume $v_{0}=0$ and $\sigma_{0}=0$ in Eq.~\eqref{eq:constantshearstress} to highlight the role of activity, rather than externally applied mechanical work, and we will separately investigate the case in which dissipation is solely hyperviscous (i.e. $\eta_{2}=0$, Sec.~\ref{sec:eta0=0}) and both viscous and hyperviscous (i.e. $\eta_{2} \ne 0$, Sec.~\ref{sec:general}). Furthermore, to set aside any undesired complexity and isolate the essential physical mechanisms at play, we hereafter set $\eta_{4}=0$
and focus on the ideal case of an infinite channel with periodic boundaries, so that $\theta(0,t)=\theta(L,t)$ and $v_{x}(0,t)=v_{x}(L,t)$. In practice, this implies restricting an infinitesimal perturbation of the quiescent state to be periodic in space, so that its wave number is an integer multiple of $q_{\min}=2\pi/L$.

In the following, we will see how decreasing $q_{\min}$ -- i.e. by progressively increasing the size of the width of the channel -- and increasing the Deborah number $\De$ -- i.e. by increasing activity or decreasing the rate of momentum dissipation -- results into a hierarchy of hydrodynamic instabilities, which allows our model epithelial layer self-organizing in an arbitrary number of oppositely flowing lanes. 

\subsection{\label{sec:eta0=0}Pure hyperviscous dissipation}

We start our analysis from the simple case in which $\eta_2=0$ so that
\begin{equation}\label{eq:viscous_simplified}
\sigma_{xy}^{({\rm v})} = \eta_{6}\partial_{y}^{4}(\partial_{y}v_{x})\;,
\end{equation}
which is sufficient to resolve unphysical effects caused by small scale instability. In this case, and for $\alpha_{6}>0$ and $\theta_{0}=0$, Eq.~\eqref{eq:groutheta0eta1eta2} demands the amplitude of linear perturbation to decrease monotonically with the wave number and change in sign at a specific activity-dependent wave number $q_{+}$, as shown by the red curve in Fig.~\ref{fig:dispersion}. The latter implies that, for small activities and narrow channels, that is when $q_{\min}>q_{+}$, the quiescent state is linearly stable to infinitesimal perturbations. By contrast, increasing activity or the channel's width, that is for $q_{\min}<q_{+}$, the stationary and uniformly oriented configuration becomes unstable to a state of spontaneous distortion and flow.

\begin{figure}
\centering
\includegraphics[width=\columnwidth]{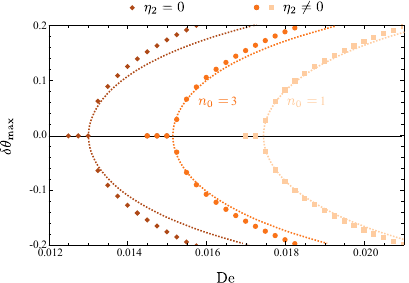}
\caption{\label{fig:instability}
\textbf{Effect of activity on the amplitude and critical mode of the stationary states} (a) Maximal angular distortion $\delta\theta_{\max}=\max_{\theta}(\theta-\theta_{0})$ obtained from a numerical integration of Eqs.~\eqref{eq:homogeneouseta0=0} and \eqref{eq:dynamicaleqchannel} for for different cases: $\eta_2=0$, $\eta_6\neq0$ (brown diamonds), $\eta_2\neq0$, $\eta_6\neq0$ and $n_0=3$ (orange dots) and $\eta_2\neq0$, $\eta_6\neq0$ and $n_0=1$ (beige squares). All curves are obtained as stationary solutions of Eqs.~\eqref{eq:generalsysteminstability}.}
\end{figure}

To gain insight into this instability and the post-transitional scenario, we use Eq.~\eqref{eq:viscous_simplified} to recast  Eqs.~\eqref{eq:constantshearstress} in the form of a single stationary equation: i.e.
\begin{equation}\label{eq:homogeneouseta0=0}
L^{4}\partial_{y}^{4}\left(L^{2}\partial_y^2\theta+\frac{\De}{16}\,\sin{6\theta}\right) = 0\;,
\end{equation}
with $\De=|\alpha_{6}|L^{2}/(\eta_{6}\mathcal{D}_{6})$ the Deborah number, as defined in Eqs.~\eqref{eq:re_and_de} for $p=6$, at the length scale of the channel width. Next, taking $\theta(y)=\theta_{0}+\delta\theta(y)$ and expanding the left-hand side at the linear order in $\delta\theta$ gives
\begin{equation}\label{eq:linearizedhomogeneouseta0=0}
\partial_{y}^{4}\left(\partial_y^2+q_{+}^{2}\right)\delta\theta=0\;,
\end{equation}
where, for simplicity, we have set $\theta_{0}=0$ and 
\begin{equation}\label{eq:critical_de_L}
q_{+} = \frac{\sqrt{\frac{3}{8}\,\De}}{L}\;.	
\end{equation}
Eq.~\eqref{eq:linearizedhomogeneouseta0=0} admits now a simple solution of the form
\begin{equation}\label{eq:linear_solution}
\delta\theta(y) = \epsilon \cos (q_{+}y+\phi)\;, 
\end{equation}
where $\epsilon$ and $\phi$ real constants. Now, for $q_{\min}>q_{+}$, that is for $\De<\De^{*}=32\pi^{2}/3$, Eq.~\eqref{eq:linear_solution} does not satisfy the periodic boundary conditions unless $\epsilon=0$, so that $\theta=\theta_{0}=0$. As soon as $\De=\De^{*}$, on the other hand, $q_{\min}=q_{+}$ and the channel is sufficiently wide to accommodate a full wavelength of the distorted configuration. As a consequence, the quiescent state becomes linearly unstable to a state of collective migration. In proximity of the transition, the configuration of the local orientation $\delta\theta$ and velocity $\delta v_{x}$ can be obtained upon expanding Eq.~\eqref{eq:homogeneouseta0=0} up to third order in $\delta\theta$. Then, taking $\phi=0$ in Eq.~\eqref{eq:linear_solution} and isolating $\epsilon$ at $y=0$, one finds 
\begin{subequations}\label{eq:completesolsimplifiedpbc}
\begin{gather}
\delta\theta(y)\approx\pm \sqrt{\frac{\De-\De^{*}}{6\De}}\,\cos(q_{+}y)\;,\\
\delta v_x(y)\approx \mp \frac{4\pi\mathcal{D}_{6}}{L}\,\sqrt{\frac{\De-\De^{*}}{6\De}}\,\sin(q_{+}y)\;.
\end{gather}	
\end{subequations}
In Eq.~(\ref{eq:completesolsimplifiedpbc}b), the configuration of the velocity field has been obtained by imposing a vanishing net velocity: i.e. $\int {\rm d}y\,v_{x}(y)=0$. A comparison between Eqs.~\eqref{eq:completesolsimplifiedpbc} and a numerical solution of Eq.~\eqref{eq:homogeneouseta0=0} is shown in Fig.~\ref{fig:instability}.

This instability is analogous to the spontaneous flow transition investigated by Voituriez {\em et al}. in active nematics~\cite{Voituriez:2005} and, because of the generality of its derivation, provides further insight on how the injection and dissipation of energy in active liquid crystals conspire to give rise to a non-equilibrium steady state, where the system spontaneously partitions in two ``lanes'' of cells flowing in opposite directions along the channel. Consistently with our discussion in Sec.~\ref{sec:regularization}, the latter example of dynamical self-organization requires energy to be injected and dissipated at the same rate across length scales and, in particular, at the largest length scale available in the channel -- i.e. the channel's width $L$ -- setting the wave length of the lowest energy mode of the orientational perturbation $\delta\theta$, that is the first one to be excited.

. 

\subsection{\label{sec:general}General case}

A more general and possibly more realistic case involves a viscous stress with finite $\eta_{2}$ and $\eta_{6}$ values. Kinetic energy is now dissipated at a rate that grows like $q^{6}$ at small length scales and like $q^{2}$ at large length scales, while these two dissipation regimes crossover at the length scale 
\begin{equation}
\ell_{\times} = \left(\frac{\eta_{6}}{\eta_{2}}\right)^{\frac{1}{4}}\;.
\end{equation}
Crucially, in circumstances that we will next discuss, this additional length scale can replace the channel width $L$ in dictating the hydrodynamic stability of the cellular fluid, as well as the size of the coherent structures emerging in the post-transitional scenario. To illustrate this concept, we start by noticing that the black curve in Fig.~\ref{fig:dispersion}, corresponding to the dispersion relation of Eq.~\eqref{eq:groutheta0eta1eta2} with $\eta_{4}=0$ and finite $\eta_{2}$ and $\eta_{6}$ values, attains negative $\omega$ values only in the range $q<q_{-}$ and $q>q_{+}$, while modes with $q_{-}<q<q_{+}$ are unstable to spatial variations. The zeros $q_{\pm}$ can be readily found from Eq.~\eqref{eq:groutheta0eta1eta2}. That is
\begin{equation}\label{eq:expressionq+-}
q_\pm=\frac{\sqrt{\frac{3}{16}\,\De_{\times}\pm\sqrt{\left(\frac{3}{16}\,\De_{\times}\right)^2-1}}}{\ell_{\times}}\;.
\end{equation}
where the quantity
\begin{equation}
\De_{\times} = \frac{|\alpha_{6}|\ell_{\times}^{2}}{\eta_{6}\mathcal{D}_{6}}=\frac{|\alpha_{6}|}{\sqrt{\eta_{2}\eta_{6}}\,\mathcal{D}_{6}}\;,	
\end{equation}
denotes again the Deborah number, but now expressed at the crossover length scale $\ell_{\times}$, rather than the channel width $L$. Now, when $\De_{\times}<16/3$, $q_{\pm}$ are complex and the quiescent state of the hexatic epithelial layer is {\em always} hydrodynamically stable, regardless of the channel width. Conversely, for $\De_{\times}>16/3$, $q_{\pm}$ are real and the quiescent state can become unstable to a state of collective migration, provided the channel is sufficiently large. 

\begin{figure}[t!]
\centering
\includegraphics[width=\columnwidth]{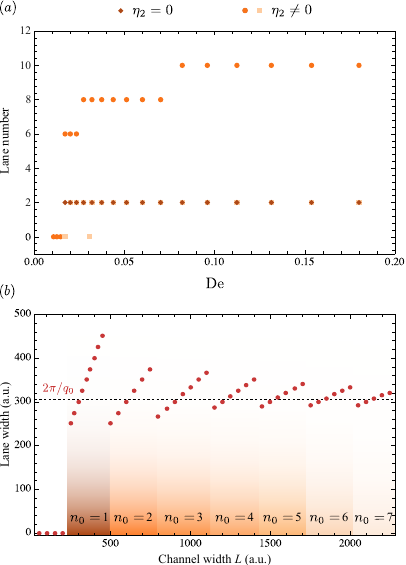}
\caption{\label{fig:lanes}
\textbf{Pattern formation under channel confinement.} (a) Number of counter-flowing lanes versus Deborah number for the same parameter values as in Fig.~\ref{fig:instability}. In the presence of both viscous and hyperviscous dissipation (i.e. $\eta_{2}\ne 0$), the system partitions in $n_{0}=2 \lfloor q_{0}/q_{\min} \rfloor$ lanes, provided $\De_{\times}>16/3$. (b) When $q_{0}<q_{\min}<q_{+}$, $n_{0}=1$ and each lane occupies half of the channel width. Increasing the channel width causes the number of lanes to increase and their width and their width to converge towards the scale-independent limit $2\pi/q_{0} \sim \ell_{\times}$ (dashed line).}
\end{figure}

This ability of controlling hydrodynamic stability through viscosity, independently on the system size, is a remarkable feature of hexatic epithelia, which, as in Refs.~\cite{Armengol:2023,Armengol:2024}, highlights the profound connection between $p-$atic order and {\em multiscale} organization in active liquid crystals. This feature appears especially powerful in the context of epithelial layers, where shear viscosity is believed to be associated with the cadherin-mediated lateral interactions between cells. Thus, increasing lateral dissipation -- i.e. by incrementing $\eta_{2}$, $\eta_{6}$ or both -- results in a stabilization of the quiescent state {\em irrespective} of channel width. In epithelia, this can be achieved by increasing the expression of E-cadherin, that is the hallmark epithelial phenotypes along the epithelial-mesenchymal spectrum.

As in Sec.~\ref{sec:eta0=0}, a deeper understanding of the post-transitional scenario can be achieved by casting Eqs.~\eqref{eq:constantshearstress} in the form of a single differential equation. This gives
\begin{equation}\label{eq:dynamicaleqchannel}
L^{4}\partial_{y}^{4}\left(L^{2}\partial_{y}^{2}\theta+\frac{\De}{16}\,\sin 6\theta\right)+\frac{L^{6}}{\ell_{\times}^{4}}\,\partial_{y}^{2}\theta = 0\;.
\end{equation}
Then, taking again $\theta(y)=\theta_0+\delta\theta(y)$, linearizing about $\theta_0=0$ and solving the resulting equation with periodic boundary conditions give
\begin{equation}\label{eq:gensolpbc}
\delta\theta(y)=\epsilon_+\cos{(q_+y)}+\epsilon_-\cos{(q_-y)}\;,
\end{equation}
with $\epsilon_{\pm}$ constants and $q_{\pm}$ as given in Eq.~\eqref{eq:expressionq+-}. When $q_{\pm}$ are real-valued and $q_{\min}>q_{\pm}$, the quiescent steady state is stable to infinitesimal perturbations. Decreasing $q_{\min}$ eventually drives a spontaneous flow transition such as that discussed in Sec.~\ref{sec:eta0=0}. In this case, however, the transition is followed by a hierarchy of additional pattern-forming instabilities originating at an intermediate length scale, corresponding to the maximum of the black curve in Fig.~\ref{fig:dispersion}a and whose specific wave number $q_{0}$ is obtained upon differentiating Eq.~\eqref{eq:groutheta0eta1eta2}. This yields the following algebraic equation 
\begin{equation}\label{eq:qcritfullexpression}
\frac{\left[(\ell_{\times}q_{0})^{4}+1\right]^{2}}{(\ell_{\times}q_{0})^{2}}= \frac{3}{4}\De_{\times}\;.
\end{equation} 
Thus, for $q_{0}<q_{\min}<q_{+}$, the channel is sufficiently large to accommodate an entire wave length of the perturbation $\delta\theta$ and the epithelial layer partitions in two lanes of cells flowing in opposite directions (see Fig.~\ref{fig:instability}). For $q_{\min}<q_{0}$, on the other hand, the system further partitions in $n_{0}=2\lfloor q_{0}/q_{\min}\rfloor$ lanes (Fig.~\ref{fig:lanes}a), whose width is now proportional to $\ell_{\times}$ and independent on $L$ in the limit $L\gg\ell_{\times}$ (Fig.~\ref{fig:lanes}b). Both these properties of the lanes pattern appear sufficiently simple to be well suited to be tested in {\em in vitro} experiments.

Repeating the analysis of Sec.~\ref{sec:eta0=0} one obtains again Eqs.~\eqref{eq:completesolsimplifiedpbc}, but the critical Deborah number is in this case given by 
\begin{equation}\label{eq:critical_de}
\De^{*}=\frac{32\pi^{2}n_{0}^{2}}{3}+\frac{2}{3\pi^{2}n_{0}^{2}}\left(\frac{L}{\ell_{\times}}\right)^{4}\;.
\end{equation}
Consistently with the previous discussion, reducing the crossover length scale $\ell_{\times}$ has the effect of stabilizing the quiescent state by shifting the spontaneous flow transition towards larger and larger $\De$ value, until, in the limit $\ell_{\times} \to 0$, $\De^{*}\to\infty$ and the transition is completely suppressed. Finally, while the specific magnitude of $q_{0}$ requires solving Eq.~\eqref{eq:qcritfullexpression}, a simpler expression can be found when $\De_{\times} \gtrsim 16/3$. In this case, the maximum $q_{0}$ and the zeros $q_{\pm}$ coalesce and from Eq.~\eqref{eq:expressionq+-} one finds
\begin{equation}\label{eq:q0}
q_{0} \approx \frac{\sqrt{\frac{3}{16}\De_{\times}}}{\ell_{\times}}\;.
\end{equation}
In summary, in the generic case in where both $\eta_{2}$ and $\eta_{6}$ are finite, the availability of a scale-dependent energy dissipation rate renders the hydrodynamic behavior of hexatic epithelial layers remarkably versatile and tunable. Specifically, when kinetic energy is mainly dissipated across a range of length scales comparable to that at which is injected -- i.e. corresponding to $\De_{\times}\approx 1$ -- the cellular fluid is hydrodynamically stable, irrespective of the length scale at which it is confined. Such a {\em generic stability} is in stark contrast with the {\em generic instability} found in {\em bulk} active nematics~\cite{Simha:2002}, where an infinitesimal mechanical activity is sufficient to drive a spontaneous flow. By contrast, when kinetic energy can flux to large length scales before being dissipated -- corresponding to $\De_{\times} \gg 1$ -- the quiescent state of the epithelial layer is unstable to the formation of lanes of counter-flowing cells. In sufficiently large systems, the width of the lanes is independent on the system size an approximatively equal to $\ell_{\times}$, thereby providing collectively migrating cellular hexatics with a mechanism to dissipate an arbitrarily activity by partitioning in an increasing number of lanes. In the following section, we will nevertheless see that this mechanism too becomes unstable once the system is released from its quasi-one-dimensional spatial confinement. 

\section{Two-dimensional instabilities}\label{sec:2D}

\begin{figure}
	\centering
	\includegraphics[width=\columnwidth]{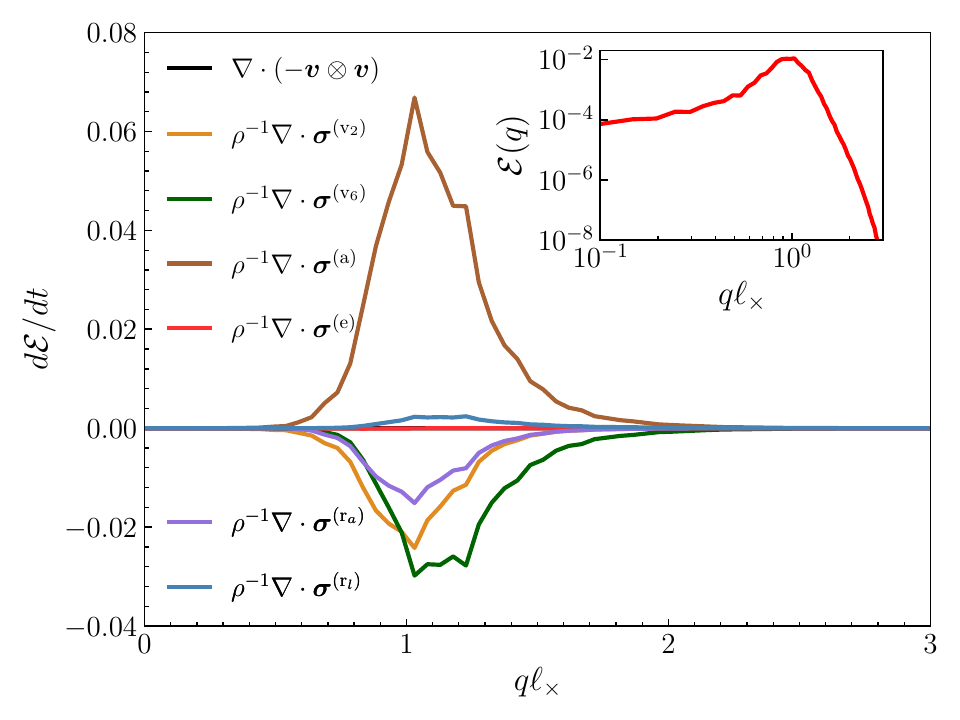}
	\caption{\label{fig:energy_budget}
\textbf{Scale by scale energy budget.} Example of the time-averaged spectral energy budget from a numerical simulation of Eqs.~\eqref{eq:hydrodynamics} in a incompressible, large activity regime. The inset shows the corresponding time-averaged kinetic energy spectrum $\mathcal{E} (q)$.}
\end{figure} 

\begin{figure}
\centering
\includegraphics[width=\columnwidth]{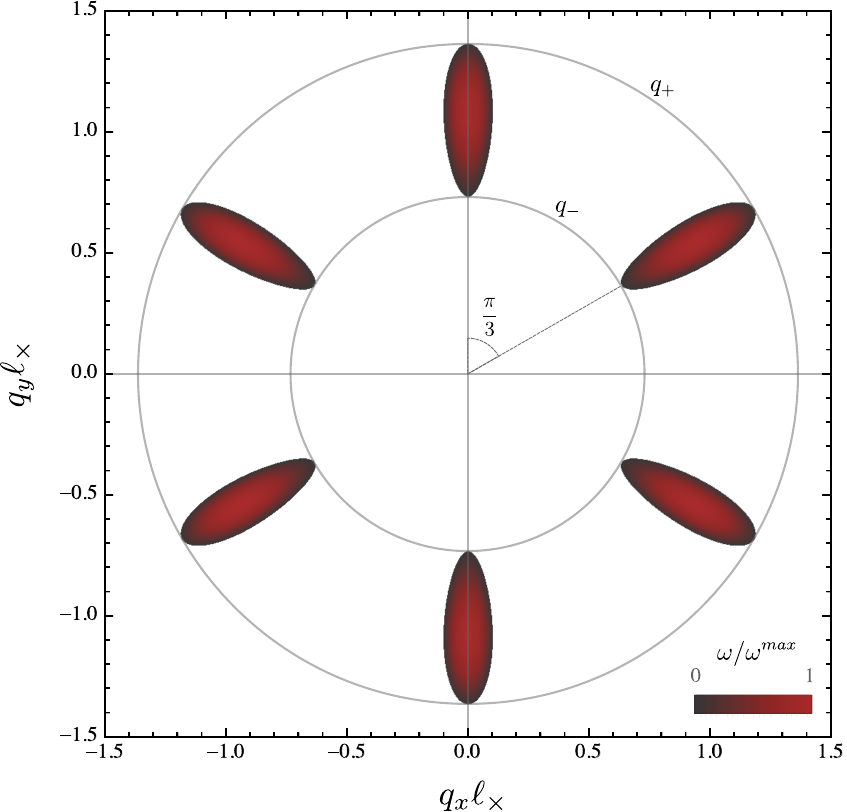}
\caption{\label{fig:2d_dispersion}
\textbf{Two-dimensional dispersion relation}. Example of the shapes defined by positive linear growth rate derived in Eq.~\eqref{eq:dispersionrelation2d} with $\De_{\times}\gtrsim 16/3$ and $\theta_0=0$. The unstable modes are delimited in the region between the concentric circles given by wavevectors with modulus $q_{-}<q<q_{+}$ and its phase is periodic with respect to rotations by $\pi/3$.}
\end{figure}

Most of the phenomenology encountered so far for the case hexatic epithelia under channel confinement, carries over to bulk monolayers. In this case, the orientation $\theta$, the flow velocity $\bm{v}$ and the pressure $P$ depends on both coordinates and must be computed by solving Eqs.~\eqref{eq:hydrodynamics} subject to the incompressibility constraint: i.e. $\nabla\cdot\bm{v}=0$. As in Sec.~\ref{sec:regularization}, we start our analysis by investigating how kinetic energy is injected or dissipated scale by scale by the various contributions to the total stress tensor. A similar analysis of different models of active fluids can be found, e.g., in Refs.~\cite{Bratanov:2015,Alert:2020,Carenza:2020a,Carenza:2020b}. At steady state and for large contractile activity, Eq.~\eqref{eq:dissipation_rate} yields the scale-dependent energy budget shown in Fig.~\ref{fig:energy_budget}, where each curve corresponds to an individual contribution to the right-hand side of the equation. We can observe how, even in a chaotic regime at large activity, the dynamics is mostly dominated by the linear viscous $\bm{\sigma}^{({\rm v})}$ and active $\bm{\sigma}^{({\rm a})}$ stresses, suggesting that important hints on the phenomenology of the system can be obtained from a linear stability analysis.

Next, taking $\theta(x,y)=\theta_{0}+\delta\theta(x,y)$ and $\bm{v}(x,y)=\bm{v}_{0}+\delta\bm{v}(x,y)$ and preceding as in Sec.~\ref{sec:periodic_boundary} yields the following linear growth rate for an angular perturbation of wave vector $\bm{q}=q(\cos\phi\,\bm{e}_{x}+\sin\phi\,\bm{e}_{y})$. That is 
\begin{equation}\label{eq:dispersionrelation2d}
\omega =- \mathcal{D}_{6}q^2-\frac{3\,\alpha_{6}q^4\cos{6(\phi-\theta_0)}}{8(\eta _2+\eta_{6}q^{4})}\;.
\end{equation}
Notice also that $\omega$ is invariant under rotations by $2\pi/6$, so the reciprocal space is now partitioned into six identical lobe-shaped regions as shown in Fig.~\ref{fig:2d_dispersion}. This symmetry considerably facilitates the stability analysis. Taking, e.g., $\phi=(\pi/6)(1+2k)$, with $k$ an integer, and solving $\Re\{\omega\}=0$ with respect to $q$ gives again the zeros $q_{\pm}$ defined in Eq.~\eqref{eq:expressionq+-}. As illustrated in Fig.~\ref{fig:2d_dispersion}, these wave numbers set the radii of the two concentric circles setting the upper and lower bounds of the regions of reciprocal space where the unstable modes are concentrated. Close to the instability, repeating the calculations of Sec.~\ref{sec:general} yields
\begin{equation}\label{eq:qcrit2d}
    \bm{q}_{0}=q_{0}\left[\sin\left(\frac{m\pi}{3}+\theta_0\right)\bm{e}_x+\cos\left(\frac{m\pi}{3}+\theta_0\right)\bm{e}_y\right]\;,
\end{equation}
where $q_{0}$ is the same as in Eq.~\eqref{eq:q0} ands $m=0,\,1\dots\,5$. 

\begin{figure}[t!]
\centering
\includegraphics[width=\columnwidth]{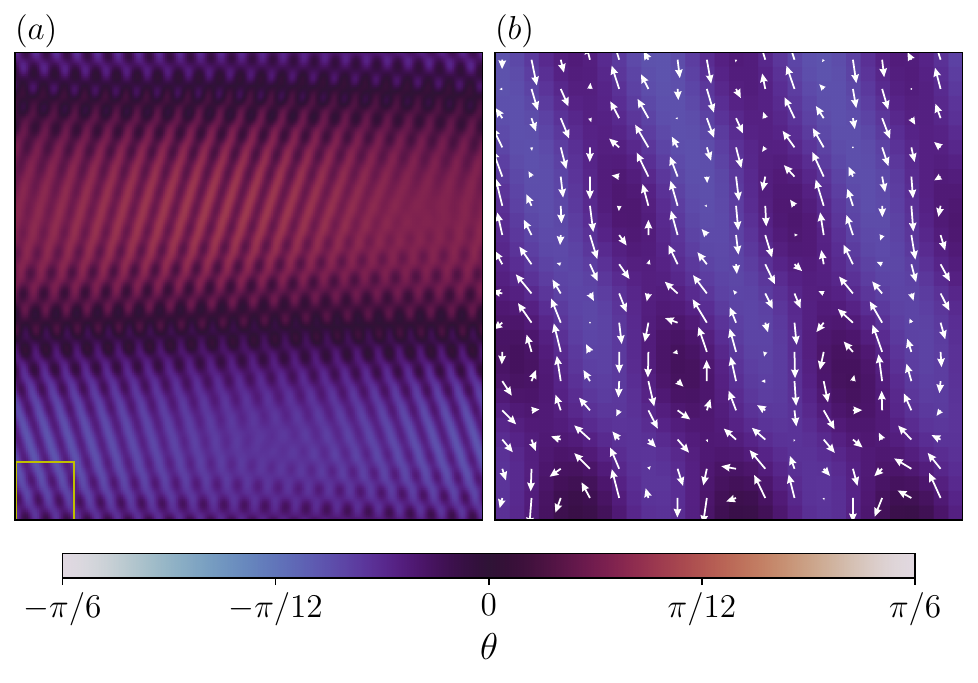}
\caption{\label{fig:bands}
\textbf{Stationary pattern at low activity.} (a) Orientation $\theta$ of the hexatic director field, characterized by ordered bands, with an orientation $\phi - \theta_0 = m\pi/6$ with respect the orientation $\theta_0 = 0$ of the initial ordered state. (b) Magnification of the yellow square in the bottom left corner of (a), with superimposed the corresponding velocity field (white arrows). The bands in the orientation correspond to laminar alternated rectilinear flows, that, in regions of intersection between pattern of different orientation, give rise to small stationary vortices.}
\end{figure}

The validity of this prediction for the general dynamics of hexatic epithelia monolayers can be tested by numerically solving Eqs.~\eqref{eq:hydrodynamics} on a two-dimensional periodic domain, using $\De_{\times}$ as control parameter and assuming incompressibility of the flow. Applying a small random perturbation on a uniformly oriented ground state at $\De_{\times} \gtrsim 16/3$, we can observe that, even re-introducing the passive backflow effects and the possibility to have $|\Psi_6| < 1$ neglected in the analytic computations, we have the  emergence of a stationary pattern (Fig.~\ref{fig:bands}), characterized by uniform bands with wavelength $2\pi/q_{0}\sim\ell_{\times}$. The orientation of the bands, on the other hand, depends on the initial conditions and, for sufficiently large systems, satisfies the condition $\phi-\theta_0=m\pi/6$. The bands are evident both in the hexatic director orientation and in the geometry of the flow, constituted by alternated rectilinear laminar flows.
	
\begin{figure}[t!]
	\centering
	\includegraphics[width=\columnwidth]{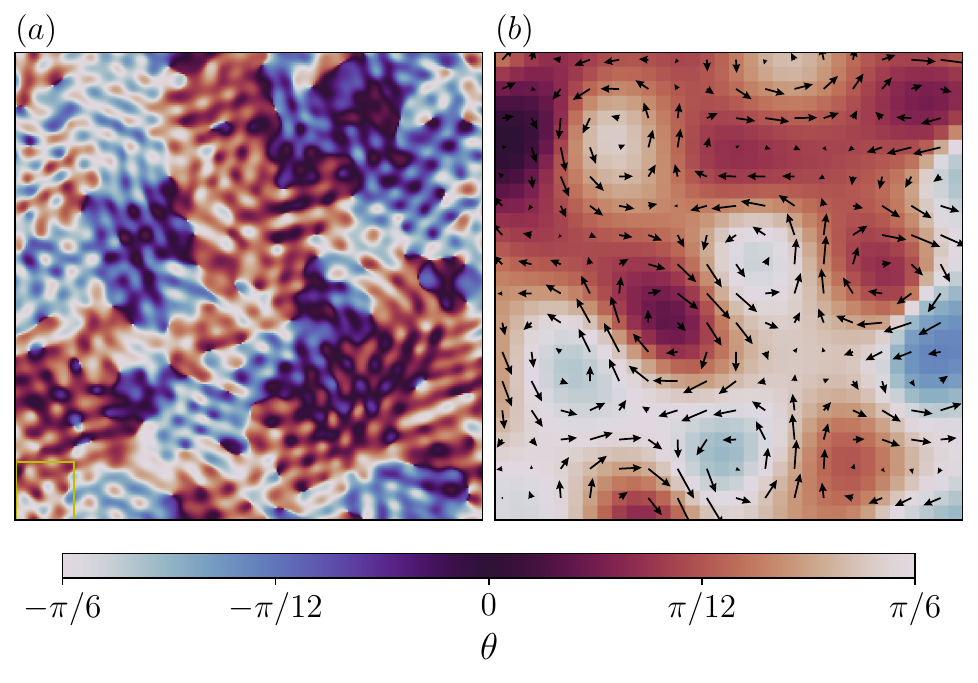}
	\caption{\label{fig:chaos}
\textbf{Spatio-temporal chaos at large activity.} (a) Orientation $\theta$ of the hexatic director, exhibiting patterns of different scales. (b) Magnification of the yellow square in the bottom left corner of (a), with superimposed the corresponding velocity field (black arrows). The disordered pattern in the orientation is reflected in the disordered motion of vortices and jets of different sizes, corresponding to the phenomenology of \textit{active turbulence.}}
\end{figure}

Increasing activity, thus $\De_{\times}$, the importance of nonlinear effects increases, and the phenomenology changes. The pattern is no longer stationary: the strips start to bend, to interact and to oscillate, giving rise to more complicated structures. The complexity of the dynamics increases with activity, eventually leading, for $\De_{\times} \gg 16/3$, towards a state characterized by spatiotemporal chaos (see Fig.~\ref{fig:chaos}), where we can clearly notice clusters of different sizes in the average orientation, along with vortical irregular structures in the flow.  The regime resulting from this further instability is reminiscent of a phenomenon observed in active nematics and usually referred to as ``active turbulence''~\cite{Giomi:2015,Alert:2020,Alert:2022}.

Nevertheless, we observe that, even in this regime, at small scale the system keeps the memory of the linear instability, as we can still observe the presence of the bands corresponding to Eq.~\eqref{eq:qcrit2d}. The scale-by-scale energy balance (Fig.~\ref{fig:energy_budget}) confirms that the nonlinear antisymmetric component of the reactive stress $\bm{\sigma}^{({\rm r})}$ become relevant in this regime and that the dynamics is mostly focused on a range of wavenumbers centered around $q_{0}$, but from the spectrum $\mathcal{E}(q)$ (inset of Fig.~\ref{fig:energy_budget}) we see that, aside from this, the flow exhibits a non-negligible amount of kinetic energy even at larger scales. 

\section{Conclusions}

In this article we have investigated the hydrodynamic stability and the formation of patterns in a continuum model of epithelial layers, with special emphasis on the role of spatial confinement. Our model, built upon recent observations of hexatic order {\em in vitro}~\cite{Armengol:2023,Eckert:2023} and {\em in silico}~\cite{Li:2018,Durand:2019,Pasupalak:2020,Li:2021}, aims at capturing the 6-fold symmetry of the stress field actively generated by the cells by virtue of their hexagonal geometry, and how the propagation of the latter affects epithelial organization at mesoscopic length scales. In this respect, the analysis presented here is not sufficiently elaborate to account for the full-fledged {\em hexanematic} behavior reported in Refs.~\cite{Armengol:2023,Eckert:2023,Armengol:2024}, but provides a simpler setting to conceptualize the essential physics originating from mechanical activity and small-scale hexatic order. Despite this simplification, the multiscale structure of epithelial layers {\em inevitably} emerges from the interplay between energy injection and dissipation, which, in epithelia as well as in any active $p-$atic liquid crystal with $p>2$, could occur with different rates at different length scales. In active hexatics, specifically, the energy injected at the scale of the cells is dissipated at the same length scale at rate $\sim \eta_{6}q^{6}$ and at larger length scales at rate $\sim \eta_{2}q^{2}$, with $\eta_{2}$ and $\eta_{6}$ generalized viscosities. The latter entails an intermediate {\em crossover} length scale, i.e. $\ell_{\times}=(\eta_{6}/\eta_{2})^{1/4}$, where these two dissipation regimes are equally prominent. 

Whether in a channel or on the plane, the existence of a non-trivial dissipative structure across length scales has two main consequence for the hydrodynamic stability and the formation of patterns in confined hexatic epithelial layers. First, for small $\ell_{\times}$ values, corresponding to strong cadherin-mediated adhesive forces among the cells, the system is {\em generically stable}. That is, the quiescent and uniformly oriented configuration of the cellular layer is stable against infinitesimal perturbations regardless of the specific length scale $L$ at which it is confined. This property is in strike contrast with the {\em generic instability} of active nematics~\cite{Simha:2002}, where any finite activity can destabilize the quiescent state, provided the system is sufficiently large. Second, when lateral adhesion is too weak to prevent active shearing, the hexatic cell layer self-organize in a hierarchy of ``lanes'' of size $\sim\ell_{\times}$. In this regime, increasing the confining length scale increments the number of lanes (i.e. $\sim L/\ell_{\times}$), without affecting their width. When confined in infinite channels, where the system is translational invariant along the channel's longitudinal direction, this hierarchy of pattern forming instabilities continue holding for arbitrarily large activities and system sizes. By contrast, when confined in a two-dimensional periodic domain, the presence of vortices compromises the hydrodynamic stability of the lane structure at large activities and the dynamics of hexatic cell layers becomes eventually chaotic. Yet, the configuration of the orientation field in Fig.~\ref{fig:chaos} -- whose herringbone pattern is reminiscent of fibrosarcoma or similar tumors~\cite{McKee:2020} -- anticipates a wealth of interesting phenomena to be explored in the near future.

\acknowledgements 

This work is supported by the ERC-CoG grant HexaTissue (L.P. and L.G.) and by Netherlands Organization for Scientific Research (NWO/OCW) as part of the research program ``The active matter physics of collective metastasis'' with project number Science-XL 2019.022 (J.-M.A.-C.). Part of this work was carried out on the Dutch national e-infrastructure with the support of SURF through the Grant 2021.028 for computational time.

\appendix

\section{Numerical methods}

Eqs.~\eqref{eq:generalsysteminstability} are solved using the open-source framework \textit{Dedalusv3}~\cite{Burns:2020}, which is specialized on the integration of spectrally-representable domains with the possibility of adding boundary conditions. More precisely, we use \textit{Fourier} and \textit{Chebyshev} basis to implement periodic and finite boundaries respectively. The specific conditions on the walls are imposed using the \textit{tau method}, which consists of adding degrees of freedom to the problem and make it solvable over polynomials.
 
The numerical integration of the complete incompressible model (Eqs~\eqref{eq:hydrodynamics} with $\nabla \cdot {\bm u}= 0$), as shown in Figs.~\ref{fig:energy_budget}, \ref{fig:bands} and \ref{fig:chaos}, has been performed with an in-house pseudo-spectral code, employing Fourier basis in order to implement periodicity along both of the spatial directions. A 4-th order Runge-Kutta scheme, with implicit integration of the viscous and diffusive terms, has been adopted, along with a velocity-vorticity formulation of Eq.~(\ref{eq:hydrodynamics}b) in order to assure the incompressibility of the flow.

\end{document}